\documentclass[12pt,a4paper]{article}

\newcommand{\be}{\begin{equation}}
\newcommand{\ee}{\end{equation}}
\newcommand{\bea}{\begin{eqnarray}}
\newcommand{\eea}{\end{eqnarray}}

\begin{document}

\title{Dimensional Reduction for Generalized Poisson Brackets}

\author{Ciprian Sorin Acatrinei\thanks{On leave from: {\it National Institute of
        Nuclear Physics and Engineering -
        P.O. Box MG-6, 077125 Bucharest, Romania}; e-mail:
        acatrine@th.if.uj.edu.pl}  \\
Smoluchowski Institute of Physics, Jagellonian University \\
Reymonta 4, 30-059, Cracow, Poland \\       
  \\  }
\date{March 21, 2007}

\maketitle

\begin{abstract}
We discuss dimensional reduction for Hamiltonian systems which possess
nonconstant Poisson brackets between pairs of coordinates and between pairs of momenta. 
The associated Jacobi identities imply that the dimensionally reduced brackets are always {\it constant}.
Some examples are given alongside the general theory.

\end{abstract}

\section{Introduction}

The dynamics of a classical conservative system with $n$ degrees of freedom is specified
once its Hamiltonian $H$ and symplectic two-from $\omega$ are given
as functions of the phase space variables $q_i$, $p_i$, $i=1,\dots,n$.
In most situations one mainly discusses specific Hamiltonians,
assuming the symplectic form to be in canonical (constant and skew-diagonal) form
\be
\omega_0=\sum_{i,j=1}^{n} dq_i\wedge dp_i.
\label{diag}
\ee 
In a context to be specified below,
we will study instead the influence on the dynamics of a generic symplectic form
\be
\omega=\sum_{a,b=1}^{2n} \omega^{ab}(x) dx_a\wedge dx_b.\label{nondiag}
\ee
Throughout this work the generalized $2n$ coordinates and momenta 
$\{q_i, p_i\}$ are occasionally denoted by $\{x_a, a=1,\dots,2n\}$. 
Darboux's theorem ensures that one can bring
the generic form (\ref{nondiag}) into the canonical form (\ref{diag}), at least locally.
However, even when the associated phase-space transformations can be explicitely performed,
the Hamiltonian becomes more complicated in the new variables.
For our purposes it will be more advantageous to work with the form (\ref{nondiag}),
as it allows a transparent implementation of the Jacobi constraints.

This paper studies the limit in which a nonconstant symplectic form $\omega$ becomes 
singular (and degenerate, taking the form
$\omega_0'\sim \sum_{i,j=1}^{n'} dq_i\wedge dp_i, n'<n$ in Darboux coordinates).
In this case the effect of the symplectic structure is maximal,
and interesting results can be obtained for generic Hamiltonians.
Consider $\Theta_{ab}$ to be the inverse of the $2n\times 2n$ matrix $\omega^{ab}$; it provides the 
Poisson brackets of the theory, $\{x_a,x_b\}=\Theta_{ab}$. The singular limit to be studied
is the one in which  the determinant of $\Theta_{ab}$ goes to zero, $\det \Theta\rightarrow 0$. 
In this limit $\omega\equiv \Theta^{-1}$ becomes singular,
and the system experiences a dimensional reduction, 
as not all the $x_a$'s have independent evolution anymore. 
The reduced system displays a regular symplectic form which, surprisingly,
turns out to be {\it constant}.
The present work is organized around the formulation and  proof of the above statement.
In this way we generalize an alternative to Peierls dimensional reduction \cite{peierls,peierls1},
alternative which was up to now discussed only for constant $\Theta_{ab}$ \cite{peierls2}. 
A discussion of the limits of validity of our results is also given. 
A more explicit approach to the singular limit 
concludes the paper.
The analysis is performed at the classical level. It can in principle be extended
to quantum mechanics if a practical ordering prescription for functions of the noncommuting operators is given.

\section{Constant symplectic form}

Let us first review the constant $\omega$ case \cite{peierls2}. Consider
\be
\omega=\sum_{a,b=1}^{2n} \omega^{ab} dx_a\wedge dx_b,\label{constant}
\ee
with the antisymetric matrix $\omega^{ab}$ having constant entries.
In the classical theory the inverse of $\omega^{ab}$, $\Theta_{ab}=(\omega^{-1})_{ab}$,
generates the extended Poisson brackets $\{x_a,x_b\}=\Theta_{ab}$.   
Quantum mechanically one replaces the Poisson brackets with commutators 
- a straightforward operation in the constant $\Theta$ case.
The formalism we discuss is applicable to any space dimensionality; 
we stay in $(2+1)$-dimensions for clarity. 
The action
\be
S=\int dt \left ( \frac{1}{2}\omega_{ab}x_a \dot{x}_b - H(x) \right ), \quad x_{1,2,3,4}=q_1,q_2,p_1,p_2,
\ee
engenders the equations of motion
\be
\dot{x}_a=\{x_a,H\}=\Theta_{ab}\frac{\partial H}{\partial x_b}, \qquad 
\Theta_{ab}=(\omega^{-1})_{ab}, \quad a,b=1,2,3,4. \label{EM}
\ee
Above the Poisson brackets are defined by  $\{A,B\}=\Theta_{ab} \partial_a A \partial_b B$, 
in particular $\{x_a,x_b\}=\Theta_{ab}$.
If we choose the symplectic form to be
\be
\Theta=
\left (
\begin{array}{rrrr}
0 & \theta & 1 & 0  \\
-\theta & 0 & 0 &1 \\
-1 &  0 & 0 & F \\
0 & -1  & -F &  0
\end{array} 
\right )
\quad \mbox{or} \quad 
\omega=\frac{1}{1-\theta F}
\left (
\begin{array}{rrrr}
0 & -F & 1 & 0  \\
F & 0 & 0 &1 \\
-1 &  0 & 0 &  -\theta\\
0 & -1  &   \theta & 0
\end{array} 
\right ), \label{omega}
\ee
the Poisson brackets are
$\{q_i,p_j\}=\delta_{ij}$, $\{q_1,q_2\}=\theta$, $\{p_1,p_2\}=F$, 
and the phase-space equations of motion become
\be
\dot{q}_i=
\frac{\partial H}{\partial p_i}+\theta \epsilon_{ij}\frac{\partial H}{\partial q_j},
\qquad
\dot{p}_i=
-\frac{\partial H}{\partial q_i}+F \epsilon_{ij}\frac{\partial H}{\partial p_j},
\qquad
\epsilon_{12}=-\epsilon_{21}=1.
\label{emh}
\ee 
Dimensional reduction occurs when $\theta F=1$. Then, 
the number of dynamical degrees of freedom is halved, since
the equations of motion imply
\be
\dot{q}_1=-\theta \dot{p}_2 \quad \mbox{and} \quad \dot{q}_2=\theta \dot{p}_1.
\label{reduction}
\ee
The degeneracy is a consequence of $\Theta$ being singular. 
The four-dimensional phase space $\{q_1,q_2,p_1,p_2\}$ collapses to a
bidimensional one,
spanned for instance by the now canonically conjugated variables $q_1$ and $q_2$.
As $\theta$ was taken to be constant, one has the
identifications
\be
q_1=-\theta p_2+c_1, \quad  \quad q_2=\theta p_1+c_2. \label{identification}
\ee
Freedom in imposing the initial conditions is insured by 
the arbitrariness of $c_1$ and $c_2$.

In $(2+1)$-dimensions, if the original Hamiltonian is rotationally invariant, 
$
H=H(p_1^2+p_2^2, q_1^2+q_2^2),
$ 
the resulting system is not only integrable - as any one-dimensional system is -
but also easily solvable. The new Hamiltonian reads
\be
H=H\left [\theta^2(q_1^2+q_2^2), q_1^2+q_2^2\right ], \quad \{q_1,q_2\}=\theta. 
\ee
In holomorphic coordinates
\be
a=\frac{1}{\sqrt{2\theta}}(q_1+iq_2), \qquad
\bar{a}=\frac{1}{\sqrt{2\theta}}(q_1-iq_2), \qquad
2\theta(q_1^2+q_2^2)=\bar{a}a,
\ee
the equations of motion read 
\be
\frac{\dot{a}}{a}=-\frac{\dot{\bar{a}}}{\bar{a}}=i\frac{dh_{\theta}}{dn},
\quad \quad
h_{\theta}(n)=h_{\theta}(\bar{a}a)\equiv H,
\ee
and are solved in terms of trigonometric functions.
At the quantum level immediate solvability is clear,
as the dimensionally reduced Hamiltonian is a function $h_{\theta}$ 
of the harmonic oscillator Hamiltonian; the spectrum is discrete, 
with energy levels given by $E_n=h_{\theta}[\theta\times(n+1/2)]$.
The simplicity of the quantum reduced Hamiltonian does not seem to be widely appreciated;
in the recent literature it was used in \cite{bns}. 

\section{Planar case}

As already stated, we will consider in this paper the 
more general case in which $\Theta$ in Eqs. (\ref{EM},\ref{omega}) is $x-$dependent.
Consider an arbitrary Hamiltonian $H(p,q)$ in (2+1)-dimensions and the 
following fundamental Poisson brackets
\be
\{q_1,q_2\}=\theta(q,p), \qquad \{p_1,p_2\}=F(q,p),\qquad \{q_i,p_j\}=\delta_{ij}.
\label{interest_2D}
\ee 
The associated Jacobi identities read
\be
\frac{\partial \theta}{\partial q_1}-F\frac{\partial \theta}{\partial p_2}=0, \quad
\frac{\partial F}{\partial p_2}-\theta \frac{\partial F}{\partial q_1}=0,
\label{jacobi1}
\ee
\be
\frac{\partial \theta}{\partial q_2}+F\frac{\partial \theta}{\partial p_1}=0,
 \quad
\frac{\partial F}{\partial p_1}+\theta \frac{\partial F}{\partial q_2}=0.
\label{jacobi2}
\ee
The  equations of motion are identical in form to the ones written in (\ref{EM}),
although $\Theta$ is not constant anymore.
They combine to yield:
\be
\dot{q}_1+\theta\dot{p}_2=(1-F\theta)\frac{\partial H}{\partial p_1}, \quad
\dot{p}_2+F\dot{q}_1=-(1-F\theta)\frac{\partial H}{\partial q_2},
\ee
\be
\dot{q}_2-\theta\dot{p}_1=(1-F\theta)\frac{\partial H}{\partial p_2},\quad
\dot{p}_1-F\dot{q}_2=-(1-F\theta)\frac{\partial H}{\partial q_1}. 
\ee
The above equations get halved in number if 
\be
F\theta=1,
\label{dr}
\ee
becoming
\be
\dot{q}_1+\theta\dot{p}_2=0, \qquad \dot{q}_2-\theta\dot{p}_1=0. \label{dimred1}
\ee
A remarkable fact, initially noted in \cite{a2}, is that one has the following

{\bf Lemma 1}  If $F\theta=1$ the Hamiltonian system
described by an arbitrary $H(p,q)$,
the brackets (\ref{interest_2D}) and the associated constraints (\ref{jacobi1},\ref{jacobi2}),
does satisfy
\be
\dot{\theta}=0.  \label{simplification}
\ee

To prove the lemma one uses the Jacobi identities (\ref{jacobi1},\ref{jacobi2}) in order to rewrite
$\dot{\theta}=\frac{\partial \theta}{\partial q_i}\dot{q}_i
+\frac{\partial \theta}{\partial p_i}\dot{p}_i$ as
\be
\dot{\theta}=\frac{\partial \theta}{\partial p_2}(\dot{p}_2+F\dot{q}_1)
+\frac{\partial \theta}{\partial p_1}(-\dot{p}_1+F\dot{q}_2)=0.
\ee 
The equality to zero above follows from $F\theta=1$ and Eq. (\ref{dimred1}).

An equivalent proof  uses the equivalence of half of Eqs. (\ref{jacobi1},\ref{jacobi2})
to $\{F,q_i\}=0$ and $\{\theta,p_j\}=0$. 
If $\theta$ and $F$ are functions of each other however,  $\{F,q_i\}=0$ implies $\{\theta,q_i\}=0$,
whereas $\{\theta,p_j\}=0$ leads to $\{F,p_j\}=0$. All the Poisson brackets of $F$ (or $\theta$) vanish
and in consequence
$$
\dot{F}=\{F,x_a\}\partial_{x_a} H=0, \qquad \dot{\theta}=\{\theta,x_a\}\partial_{x_a} H, \qquad \forall H(p,q).
$$
In brief, the requirement $\det \Theta =0$ and the Jacobi constraints suffice to
enforce Eq.(\ref{simplification}), even for a nonconstant symplectic form. 
The equations of motion were used in the proof, though for arbitrary Hamiltonian.
Eqs. (\ref{dimred1}) and (\ref{simplification}) permit again the identification (\ref{identification}),
showing explicitely that only two out of the four initial $x_a$ are independent - dimensional reduction takes place.

To show that $\theta$ is actually a trivial constant, not an integral of the motion,
the above proof can be refined by explicitely solving the Jacobi constraints.
If $F\theta=1$, their number  gets halved,
as Eqs. (\ref{jacobi1},\ref{jacobi2}) reduce to
\be
\theta\frac{\partial \theta}{\partial q_1}-\frac{\partial \theta}{\partial p_2}=0, \quad \qquad
\theta\frac{\partial \theta}{\partial q_2}+\frac{\partial \theta}{\partial p_1}=0.
\label{jacobi_red}
\ee
Using the method of characteristics for partial differential equations of first order,
the general solution $\theta(q_1,p_2,p_1,p_2)$ of the system (\ref{jacobi_red})
is found to be given implicitely by 
\be
\theta=\phi(q_1+\theta p_2,q_2-\theta p_1),
\label{solution}
\ee
$\phi$ being an arbitrary function of two independent variables.
Since (\ref{dimred1}) and (\ref{simplification}) imply again that
$$
q_1+\theta p_2=c_1, \qquad q_2-\theta p_1=c_2, 
$$  
Eq. (\ref{solution}) transforms into
\be
\theta=\phi(c_1,c_2).
\ee
Only those combinations of the $x$'s  which are becoming pure constants when $\theta F\rightarrow 1$ 
can enter $\theta$.
Consequently, $\theta^{red}$ -- the value of $\theta$ in the dimensionally reduced space --
is a pure constant itself. We thus proved

{\bf Theorem 1} Any planar dynamical system 
characterized by the brackets (\ref{interest_2D}) 
dimensionally reduces in the limit $F\theta\rightarrow 1$.
The reduced system exhibits a {\it constant} bracket between its canonically conjugated variables.
If those are chosen to be $q_1$ and $q_2$, that constant bracket is $\theta^{red}$.

As a corollary, the solvability analysis presented at the end of Section 2 remains valid for the reduced system.

We present a second proof of this theorem, from the point of view of the reduced system;
it does not use the equations of motion at all.
$\theta^{red}$ is a constant if its total variation
with respect to any initial phase space coordinate is zero, {\it after} dimensional reduction.
Consider for instance variation with respect to $q_1$. If we choose $q_1$, $q_2$
as independent (and canonically conjugated) coordinates of the reduced system, 
$p_1$ and $p_2$ will become functions of those.
The {\it total} variation of $\theta^{red}$ with respect to $q_1$ is then
\be
\frac{\delta \theta^{red}}{\delta q_1}=\frac{\partial \theta}{\partial q_1}
+\frac{\partial \theta}{\partial p_i}\frac{\partial p_i}{\partial q_1}.
\ee
It will be independently proved later - Eq. (\ref{in_fact}) - that $\frac{\partial p_k}{\partial q^n}=F_{kn}$.
Using this we obtain
\be
\frac{\delta \theta^{red}}{\delta q_1}=\frac{\partial \theta}{\partial q_1}
-F\frac{\partial \theta}{\partial p_2}=0
\label{J_red}
\ee
which vanishes due to Jacobi (\ref{jacobi1}). The same can be shown with respect to  $q_2,p_1,p_2$.
In checking this one obtains a very interesting interpretation of the Jacobi identities
from the reduced space viewpoint: they {\it precisely} ensure that the total variation with respect
to any variable is zero, cf. (\ref{J_red}).

One can examplify with a particular solution of the Jacobi constraints
which exhibits dimensional reduction, for instance $\theta^{red}=1/F^{red}=-\frac{q_1}{p_2}$.
Then $\frac{\delta \theta^{red}}{\delta q_1}\sim p_2-q_1F^{red}= 0$.
In conclusion, variation of any $x_a, a=1,2,3,4,$ does not produce - thanks to Jacobi -  a variation of 
$\theta^{red}$, which is consequently a constant 

A final consistency check can be performed: initially we started with relations of the type
$\{q_1,p_2\}=0$, which should remain true in the singular limit.
Now, if $p_2=-\frac{1}{\theta}q_1$, in the end $\{q_1,p_2\}\sim \{q_1,\theta\}$,
which vanishes in the reduced case.

\section{General result}

We proceed to the case of an arbitrary number of dimensions.
Consider first a generalized electromagnetic background $F(q,p)$, living on a space
with noncommutativity field $\theta(q,p)$, and flat metric $g_{ij}=\delta_{ij}$: 
\be
\{q_i,q_j\}=\theta_{ij}(q,p) \qquad \{q_i,p_j\}=\delta_{ij} \qquad \{p_i,p_j\}=F_{ij}(q,p).   \label{PB_FT} 
\ee
The quantum mechanical version is obtained by replacing the Poisson brackets
with commutators, $\{,\}\rightarrow -i[,]$, and requires a practical
prescription for operator ordering - this will not be adressed here.

A more elegant notation would use covariant and contravariant indices,
writing $q^k$ and $\theta^{ij}$ with the indices up.
As no risk of confusion arises in this paper, we choose to keep all the indices down for simplicity.

The Jacobi identities read
\be
\{q_k,F_{ij}\}=\frac{\partial F_{ij}}{\partial p_k}-\frac{\partial F_{ij}}{\partial q_m}\theta_{mk}=0   \label{j1}
\ee
\be
\{\theta_{ij},p_k\}=\frac{\partial \theta_{ij}}{\partial q_k}+\frac{\partial \theta_{ij}}{\partial p_m}F_{mk}=0  
\label{j2}
\ee
\be
\{F_{ij},p_k\}+cyclic=
(\frac{\partial F_{ij}}{\partial q_k}+\frac{\partial F_{ij}}{\partial p_m}F_{mk})+cyclic =0 \label{jc1}
\ee
\be
\{q_k,\theta_{ij}\}+cyclic=
(\frac{\partial \theta_{ij}}{\partial p_k}-\frac{\partial \theta_{ij}}{\partial q_m}\theta_{mk})+cyclic =0. \label{jc2}
\ee
They ensure the invariance of the commutation relations
under time evolution, for a generic Hamiltonian $H(p,q)$. Explicitely:
\be
\{\{q_m,p_n\}-\delta_{mn},H\}=-\partial_{p_s}H\{F_{ns},q_m\}+\partial_{q_s}H\{\theta_{ms},p_n\},\label{ev1}
\ee
\be
\{\{q_m,q_n\}-\theta_{mn},H\}=-\partial_{p_s}H\{\theta_{mn},p_s\}-
\partial_{q_s}H(\{\theta_{ns},q_m\}+cyclic),\label{ev2}
\ee
\be
\{\{p_m,p_n\}-F_{mn},H\}=-\partial_{q_s}H\{F_{mn},q_s\}
+\partial_{p_s}H(\{F_{mn},p_s\}+cyclic),\label{ev3}
\ee
and  Eqs. (\ref{j1}--\ref{jc2}) ensure that the right hand side of Eqs. (\ref{ev1}--\ref{ev2}) is zero.
The Jacobi identities above restrict  $F$ and $\theta$ to 
one of the following forms:
\begin{itemize}

\item $F$ and $\theta $ both constant - the simplest situation,
intensively studied recently under the name
of noncommutative quantum mechanics \cite{ncqm}. 
The dimensional reduction has been worked out previously \cite{peierls2}

\item $\theta $  constant, and $F(q,p)$  constrained by (\ref{j1},\ref{jc1}).
This case is quite interesting for a variety of reasons. 
In particular, if $\theta_{ij}\neq 0$ then Eq. (\ref{j1}) forbids an electromagnetic field strength 
to depend only on the coordinates $q$. 
A detailed study appeared in \cite{a1}. Dimensional reduction cannot occur.

\item $F$  constant, and $\theta(q,p)$  constrained by(\ref{j2},\ref{jc2}) - the dual of the above.

\item $F(q,p)$ and $\theta(q,p)$, constrained by (\ref{j1}--\ref{jc2}) - the case of interest here.

\end{itemize}

We will first show in  general  that relations of the type
"$F=\frac{1}{\theta}$" arise {\it if and only if} the $q$'s and $p$'s are not independent,
i.e. if dimensional reduction takes place. 
We work in an arbitrary number of dimensions,
with the phase space spanned by the set $\{q_i,p_j\}$.

First, assume there exists a relation between the $q_i$'s and the $p_j$'s, say,
\be
q_n=g_n(p_m), \qquad p_m=f_m(q_s)=(g^{-1})_m(q_s). \label{dependence}
\ee
Asking consistency of the commutation relations, one obtains equalities of the type
$\{f(q(p)),q_s\}=\frac{\delta f}{\delta q_m}\theta_{ms}
=-\frac{\delta f}{\delta q_m}\frac{\partial q_m}{\partial p_s}$. These imply
\be
\frac{\partial q_m}{\partial p_s}=-\theta_{ms}, \qquad 
\frac{\partial p_k}{\partial q_n}=F_{kn}.  \label{in_fact}
\ee 
Since $\frac{\partial q_m}{\partial p_k}\frac{\partial p_k}{\partial q_n}=\delta_{mn}$, it follows that 
\be
\theta_{mk}F_{kn}=-\delta_{mn}. \label{maestru}
\ee
Eq. (\ref{maestru}) actually amounts to $\det \Theta =0$, if $\Theta$ provides the Poisson brackets (\ref{PB_FT}). 
In two dimensions one immediately recovers the previous relation $F_{12}\equiv F=\theta^{-1}\equiv \theta_{12}^{-1}$.
An alternative derivation uses the following chain of equalities,
\be
\delta_{mn}=\{q_m,p_n\}=\{g_m(p),p_n\}=\frac{\partial q_m}{\partial p_s}F_{sn}=-\theta_{mk}F_{kn}.
\ee 
Eq. (\ref{maestru}) 
is consistent with the commutation relations, as illustrated by
\be
\{q_m,q_n\}=\{g_m(p),g_n(p)\}=
\frac{\partial g_m}{\partial p_s}F_{st}\frac{\partial g_n}{\partial p_t}=
\theta_{ms}F_{st}\theta_{nt}=\theta_{mn}.
\ee

Conversely, we wish to prove that the constraint (\ref{maestru}) implies dependence
of  the phase space variables (\ref{dependence}). 
Eq. (\ref{maestru}) and the equations of motion imply
\be
\dot{q}_m+\theta_{mn}\dot{p}_n=\{q_m+\theta_{mn} p_n,H\}=0, \quad \forall H(p,q).
\label{variations}
\ee
Using (\ref{maestru}), Eq. (\ref{variations}) can be put into the equivalent form
\be
\dot{p}_m-F_{mn}\dot{q}_n=0.
\label{variations_1}
\ee
Eqs. (\ref{variations},\ref{variations_1}) already show that dimensional reduction occurs,
since the variations of the $q$'s and the $p$'s are related. 
To see it more explicitely, guided by the work of the previous section, we calculate
$\dot{\theta}_{ij}$. Using the Jacobi identities we have
\be
\dot{\theta}_{ij}=\frac{\partial \theta_{ij}}{\partial x_a}\dot{x}_a=
\frac{\partial \theta_{ij}}{\partial p_m}(\dot{p}_m-F_{mn}\dot{q}_n)=0.
\ee 
The last equality made use of (\ref{variations_1}). In consequence (the proof for $\dot{F}_{ij}$ is identical)
Eq. (\ref{maestru}) and the Jacobi identities enforce
\be
\dot{\theta}_{ij}=0, \qquad \qquad \dot{F}_{ij}=0.
\label{ddot}
\ee
Eqs. (\ref{variations},\ref{variations_1}) and (\ref{ddot}) imply a precise relationship
between the $q$'s and the $p$'s: 
\be
q_m=-\theta_{mn}(q,p)p_n+c_m,   \qquad  \forall H(p,q), \label{explicit}
\ee
which generalizes the two-dimensional relation (\ref{identification}).

One may enquire if (\ref{explicit}) and (\ref{in_fact}) are compatible.
Differentiating with respect to $p$ one sees immediately that the
necessary and sufficient conditions are precisely the Jacobi identities.
Thus the whole set of constraints is consistent and we have demonstrated

{\bf Lemma 2} Given a system with $H(p,q)$ and Poisson brackets (\ref{PB_FT}),
the constraints (\ref{maestru}) and (\ref{explicit}) are equivalent. Either
of them implies dimensional reduction, and consequently leads to (\ref{ddot}).

$F(q,p)$ and $\theta(q,p)$ are thus constants of the motion. 
Since the analysis was independent of the form of the Hamiltonian, they are expected to be trivially constant.
Otherwise, if for instance $H(p)$, then $\theta(q,p)$ would be 
a "constant of the motion" unrelated to the Hamiltonian.
One way to demonstrate trivial constancy is to search for solutions of the Jacobi identities
(\ref{j1},\ref{j2},\ref{jc1},\ref{jc2}) {\it and} the constraint (\ref{maestru}). 
One first notices an important fact:
(\ref{maestru}) ensures that Eqs. (\ref{jc1},\ref{jc2}) are automatically satisfied, 
provided the first two identities, Eqs. (\ref{j1},\ref{j2}), hold.
The idea now is to show that the solutions of Eqs. (\ref{j1},\ref{j2}) 
are given by functions which depend on $\bar{q}_m\equiv q_m+\theta_{mn}p_n$ 
(or equivalently on $\bar{p}_l\equiv p_l+F_{lm}q_m)$ 
and to invoke (\ref{explicit}) - which shows that exactly those combinations are constant.
Let us sketch the proof of our claim. 
Consider $\theta_{ij}=f_{ij}(q_m+\theta_{mn}p_n)\equiv f_{ij}(\bar{q}_m)$.
Then
$$
\frac{\partial \theta_{ij}}{\partial q_k}
=\frac{\partial f_{ij}}{\partial \bar{q}_m}(\delta_{km}
+\frac{\partial \theta_{mn}}{\partial q_k} p_n), \quad \quad
\frac{\partial \theta_{ij}}{\partial p_m}
=\frac{\partial f_{ij}}{\partial \bar{q}_s}
(\frac{\partial \theta_{sn}}{\partial q_k}p_n+\theta_{sm}).
$$
Denoting $\frac{\partial f_{ij}}{\partial \bar{q}_m}p_n$ by $A_{ij;mn}$ one sees that
$\frac{\partial \theta_{ij}}{\partial q_k}$ and $\frac{\partial \theta^{ij}}{\partial p_m}$ 
are to be found from two systems of $n(n-1)/2$ linear equations each,
\be
(A_{ij;sn}-\delta_{is}\delta_{jn})\frac{\partial \theta_{sn}}{\partial q_k}=\frac{\partial f_{ij}}{\partial \bar{q}_k},
\quad \quad
(A_{ij;sn}-\delta_{is}\delta_{jn})\frac{\partial \theta_{sn}}{\partial p_m}=
\frac{\partial f_{ij}}{\partial \bar{q}_s}\theta_{sm}.
\ee
The two systems differ only through their inhomogeneous terms: the first set of inhomogeneous terms
produces the second upon contraction with $\theta_{mn}$. This immediately shows that the
corresponding solutions of the two systems are obtained from one another
through the same contraction - and this gives exactly the required Jacobi identity. 
Using the notation $\theta_{ij}^{red}$, $F_{ij}^{red}$ for the reduced system values, we have

{\bf Theorem 2} The Hamiltonian system described by an arbitrary $H(p,q)$ and by the Poisson brackets (\ref{PB_FT})
dimensionally reduces if the constraint (\ref{maestru}) is imposed. All the Poisson brackets are
pure constants in the reduced system; in particular $\theta_{ij}^{red}$ and $F_{ij}^{red}$ are all 
trivially constant. 

It is instructive to give a second, simpler proof, from the point of view of the reduced system.
It relies on one single but essential fact. The Jacobi identities, when restricted to the reduced system
in which the $x$'s are related by (\ref{explicit}),
mean that the total variation of $F^{red}$ or $\theta^{red}$ with respect to any coordinates is zero.
That the Jacobi identity $\{\theta^{mn},p_l\}=0$ means zero  total variation of $\theta$ in reduced space
can be seen from the following sequence of equalities:
\be
\{\theta_{mn},p_l\}=\frac{\partial \theta_{mn}}{\partial q_l}+F_{sl}\frac{\partial \theta_{mn}}{\partial p_s}\\
=F_{sl}^{red}\left (\frac{\partial \theta_{mn}^{red}}{\partial p_s}+
\frac{\partial \theta_{mn}^{red}}{\partial q_r}\frac{\partial q_r}{\partial p_s}\right )
=F_{sl}^{red}\frac{\delta \theta_{mn}^{red}}{\delta p_s}=\frac{\delta \theta_{mn}^{red}}{\delta q_l}
\ee
in which we used (\ref{in_fact}) repeatedly. 
Thus $\{\theta_{mn},p_l\}=0$ implies both $\frac{\delta \theta_{mn}^{red}}{\delta q_l}=0$
and $\frac{\delta \theta_{mn}^{red}}{\delta p_s}=0$.
Similarly, $\{F_{mn},q_l\}=0$ implies
$\frac{\delta F_{mn}^{red}}{\delta q_t}=F_{st}^{red}\frac{\delta F_{mn}^{red}}{\delta p_s}=0$.

Thus  $F_{ij}$ and $\theta_{mn}$ are either constant, or display (thanks to Jacobi)
only $q,p$-dependencies which lead to constancy when $q=q(p)$. 
In two-dimensions such examples are $F_{12}(p_1,q_2)=\theta^{-1}_{12}=\frac{p_1}{q_2}$; 
$F_{12}(p_2,q_1)=\theta^{-1}_{12}=-\frac{p_2}{q_1}$,
or a more general one, $F_{12}(p_1,p_2,q_1,q_2)=\theta^{-1}_{12}=\frac{p_1+p_2}{q_2-q_1}$.
It is easy to check that any of those satisfies $\{F_{12},q_l\}=0$, once $F_{12}=\theta_{12}^{-1}$.

We showed that for a large class of systems the Poisson brackets
of the reduced system are not only time-independent, but completely constant.
In doing so we considerably generalized an alternative \cite{peierls2} 
to the celebrated Peierls reduction \cite{peierls,peierls1} in strong magnetic fields.
An elementary corollary of our aproach is immediate integrability of the reduced system,
at least if the starting point is a rotationally invariant planar Hamiltonian. 
Some higher dimensional examples are currently under investigation.

\section{Other cases}

Consider the most general planar Poisson brackets
\be
\{q_1,q_2\}=\theta(q,p), \qquad \{p_1,p_2\}=F(q,p),\qquad \{q_i,p_j\}=g_{ij}(q,p)
\label{most_general_2D}
\ee
and their analogues in higher dimensions.
The brackets $g_{ij}$ became generic functions of $q_i,p_j$.
We would like to clarify in which conditions the results of Sections 3 and 4 continue to hold.

The equations of motion are of the form $\dot{x}_a=\Theta_{ab}\frac{\partial H}{\partial x_b}$
and are written down straightforwardly by use of (\ref{most_general_2D}).
In $(2+1)$- dimensions the  null-determinant condition $\det \Theta =0$ reads
\be
\theta F-g_{11}g_{22}+g_{12}g_{21}=0.
\label{det_nul}
\ee
It permits to show that four linear combinations of the time derivatives of the $x_a$'s give zero
\be
F\dot{q}_1-g_{12}\dot{p}_1+g_{11}\dot{p}_2=0, \quad \quad 
F\dot{q}_2-g_{22}\dot{p}_1+g_{21}\dot{p}_2=0,
\ee
\be
g_{22}\dot{q}_1-g_{12}\dot{q}_2+\theta\dot{p}_2=0, \quad \quad
g_{21}\dot{q}_1-g_{11}\dot{q}_2+\theta\dot{p}_1=0.
\ee
However only two of the above constraints are independent, showing that only two out of the four
time derivatives $\dot{q}_1,\dot{q}_2,\dot{p}_1,\dot{p}_2$ have independent evolution.
Dimensional reduction takes place when (\ref{det_nul}) is satisfied.

Define for future use the linear differential operators
\be
D_1=-\theta\frac{\partial}{\partial q_2}-g_{11}\frac{\partial}{\partial p_1}-g_{12}\frac{\partial}{\partial p_2},
\quad \quad 
D_2=-\theta\frac{\partial}{\partial q_1}+g_{21} \frac{\partial}{\partial p_1}+g_{22}\frac{\partial}{\partial p_2},
\ee 
\be
D_3=-F\frac{\partial}{\partial p_2}+g_{11}\frac{\partial}{\partial q_1}+g_{21}\frac{\partial}{\partial q_2},
\quad \quad 
D_4=+F\frac{\partial}{\partial p_1}+g_{12}\frac{\partial}{\partial q_1}+g_{22}\frac{\partial}{\partial q_2}.
\ee
Only two of them are linearly independent, as for instance
\be
g_{11}D_2=g_{12}D_3-\sigma D_1, \quad \quad -g_{11}D_4=\theta D_3+g_{21} D_1,
\label{4to2}
\ee
but using all four on equal footing simplifies the writing.
To calculate a time derivative $\dot{f}=\frac{\partial f}{\partial x_c}\dot{x}_c$
of a function $f(x)$ we observe
(after straightforward manipulation of the equations of motion) that
\be
\frac{d}{dt}=
\frac{\partial H}{\partial q_1}D_1-\frac{\partial H}{\partial q_2}D_2
+\frac{\partial H}{\partial p_1}D_3+\frac{\partial H}{\partial p_2}D_4.
\label{time_derivative}
\ee
To ensure that $\dot{f}=0$, it is thus enough to show 
- given Eqs. (\ref{4to2}) - that $D_1 f=0$ and $D_3 f=0$.

Additional information is given by the Jacobi identities, which read
\be
D_3\theta+D_1g_{21}+D_2g_{11}=0, \quad \quad D_4\theta+D_1g_{22}+D_2g_{12}=0,
\label{J_GEN_1}
\ee
\be
D_1 F-D_3g_{12}+D_4g_{11}=0, \quad \quad D_2 F-D_3 g_{22}+D_4 g_{21}=0.
\label{J_GEN_2}
\ee 
(Of course, the differential operators are supposed to act on all the functions which are put on their right.)
There is not enough information in (\ref{J_GEN_1},\ref{J_GEN_2}) to show that  $D_1$ and $D_3$ (say)
annihilate any of the functions $F$, $\theta$ or $g_{ij}$. We must retreat to particular cases.

If the $g_{ij}$'s are constant one obtains 
$D_1\theta=D_4\theta=0$, $D_2 F=D_3 F=0$, enough to prove their
constancy; $F$ and $\theta$ are also related via (\ref{det_nul}).

However, if we take $g_{12}=g_{21}=0$, but all the other four nonconstant, all that we can reach is
\be
D_3(\frac{\theta}{g_{11}})=D_3(\frac{g_{22}}{F})=0 
\quad \quad 
D_1(\frac{\theta}{g_{22}})=D_1(\frac{g_{11}}{\theta})=0.
\ee
which is not enough to prove anything, except that $\frac{\theta F}{g_{11}g_{22}}$ is constant. 
However this expression is already known to be equal to $1$, cf. (\ref{det_nul}).

An intermediate situation appears if three of the six functions present in (\ref{most_general_2D})
are nonconstant, e.g. $\theta$, $F$ and one $g$, say $g_{12}$. Then one can use (\ref{det_nul}) to express 
$g_{12}$ as a function of $\theta$ and $F$, which are then constrained to be constant by the four 
identities (\ref{J_GEN_1},\ref{J_GEN_2}).

We see that not only the Poisson structure (\ref{interest_2D}) has the remarkable property of
forcing its dimensionally reduced brackets to be constant. The condition $\det \Theta =0$ allows
in fact for one more nonconstant bracket in (\ref{most_general_2D}). 
Making use of the symmetry between the $x$'s, 
the three nonconstant brackets in  (\ref{interest_2D}) can be picked at will, and we have demonstrated

{\bf Theorem 3}
If at most three out of the six functions appearing in (\ref{most_general_2D}) are nonconstant
and  the dimensional reduction condition (\ref{det_nul}) is imposed, then the reduced brackets
are all constant.

It is now clear why the property of the structure (\ref{interest_2D})
generalizes to any even number of dimensions, namely for brackets of the type (\ref{PB_FT}).
In this case on has $n(n-1)$ functions, $\theta_{ij}$, $F_{ij}$, $i=1,2\dots,n$,
on which the Jacobi identities (\ref{j1},\ref{j2}) impose exactly the same number
of constraints, namely that each of them be annihilated by the operator
$\frac{\partial}{p_k}-\theta_{mk}\frac{\partial}{\partial q_m}$.
Here too one single nonconstant $g$ can be added to the game, as it is determined
by $\theta_{ij}$ and $F_{ij}$ via the $\det \Theta =0$ condition. Using again the freeedom to relabel
the $q$'s and $p$'s at will, one can prove our most general statement:

{\bf Theorem 4} Given a $2n$-dimensional phase space $\{x_a, a=1,\dots,2n\}$ with $2n^2-n$ brackets
$\Theta_{ab}=\{x_a,x_b\}$, if at most $n^2-n+1$ of them are given by nonconstant functions,
then the condition $\det \Theta =0$ forces all of them into constants.

\section{The singular limit explicitely}

In the previous sections the constraints posed by the Jacobi identities 
were solved in the dimensionally reduced (singular) case.
To explicitely see dimensional reduction at work, we wish to solve in $(2+1)$-dimensions 
the initial constraints (\ref{jacobi1},\ref{jacobi2}).
Since those can be completely separated in equations containing derivatives either
with respect to $q_1$ and $p_2$, or $q_2$ and $p_1$, we consider only the
pair $(q_1,p_2)$. Defining for simplicity $u=F$, $1/\theta=v$, $q_1=x$, $p_2=y$, 
$\partial_x=\partial_1$, $\partial_y=\partial_2$, the equations to be solved are:
\be
\partial_1 u- v\partial_2 u=0, \quad\quad \partial_1 v -u \partial_2 v=0.
\label{general_system}
\ee
We are interested in the existence of solutions with $u\neq v$ in general, 
but with $u\rightarrow v$ as a parameter is varied (a smooth approach to dimensional reduction in a sense).
We will forget for a moment about the physical dimensions of $q,p,F,\theta$,
as these can be easily reinstated at a later stage through the introduction of
dimensionfull parameters.
Since the above system is nonlinear, but with coefficients not depending on the two
independent variables, we use the hodograph method - we invert the roles of the
dependent and independent variables. 
Seeing $x$ and $y$ as functions of $u$ and $v$ is possible if the Jacobian 
of the transformation 
$$
J= \partial_x u\partial_y v-\partial_y u\partial_xv
$$
is nonzero (otherwise one easily shows that all solutions satisfy $u=v$).
Eqs. (\ref{general_system}) become
\be
\partial_u y+ u\partial_u x=0, \quad\quad \partial_v y +v \partial_v x=0.
\label{inverted}
\ee
One shows easily that the only solutions with $u\neq v$ of the above system
of linear partial differential equations are of the form
\be
x(u,v)=f(u)+g(v), \qquad \quad y(u,v)= -\int u \frac{df}{du}-\int v\frac{dg}{dv},
\label{gen_sol}
\ee 
with $f$ and $g$ arbitrary functions of one variable.
To fix $f$ and $g$ one needs boundary conditions or initial conditions
coming from a physical criterion. We do not adress this here, but merely
present a few mathematically  simple choices for $f$ and $g$.

Consider first the choice $f=\alpha u$, $g=-\alpha v$. Then one obtains in the end
\be
u=-\frac{y}{x}+\frac{x}{2\alpha}, \quad \quad v=-\frac{y}{x}-\frac{x}{2\alpha}. 
\label{exp_1}
\ee
Restoring dimensionality one sees that $\alpha$ has dimension $[length]^{3}$.
It is immediately seen that
$\lim_{\alpha \rightarrow \infty} u=\lim_{\alpha \rightarrow \infty}v= -y/x$ 
reproduces the reduced solution mentioned earlier.

Another simple choice, $f(u)=\alpha \log (u/u_0)$, $g(v)=-\alpha \log (v/u_0)$,  leads to 
\be
u=\frac{(y/\alpha) e^{x/\alpha}}{1-e^{x/\alpha}}, \quad \quad v=\frac{(y/\alpha)}{1-e^{x/\alpha}}.
\label{exp_2}
\ee
This time $\alpha$ has the physical dimension of $[length]^{1}$ and $[u_0]=[length]^{-2}$.
The limiting behaviour is the same as in the first example, $u=v=-\frac{y}{x}$
or, in initial notation,  $F=1/\theta=-\frac{p_2}{q_1}$.

An apparently innocent change of the above can make a big difference. 
The choice $f(u)=\alpha \log (u/u_0)$, $g(v)=\alpha \log (v/v_0)$, with $\alpha$ a length
and $u_0,v_0$ having dimension $[length]^{-2}$, leads to
$$
u=\frac{-y/\alpha+\sqrt{y^2/\alpha^2-4u_0v_0e^{x/\alpha}}}{2}, \quad  
v=\frac{-y/\alpha-\sqrt{y^2/\alpha^2-4u_0v_0e^{x/\alpha}}}{2},
\label{exp_3}
$$
or viceversa.
This is an example of solution which does not display dimensional reduction ($u=v$) in any limit.

\bigskip

{\bf Acknowledgements}

During its various stages this work was supported by the Marie Curie Actions 
Transfer of Knowledge Project COCOS (contract MTKD-CT-2004-517186),
by NATO Grant PST.EAP.RIG.981202, and by Romanian Grant CEEX-05-D11-49.


\end{document}